\documentclass[letterpaper,english,reprint,aps,prl,superscriptaddress,showpacs]{revtex4-1}
\usepackage[latin9]{inputenc}
\usepackage{amsmath}
\usepackage{amssymb}
\usepackage{graphicx}

\makeatletter

\pdfpageheight\paperheight
\pdfpagewidth\paperwidth

\newcommand{\lyxdot}{.}

\usepackage{babel}
\usepackage[colorlinks,
            linkcolor=red,
            anchorcolor=black,
            citecolor=blue
            ]{hyperref}

\makeatother

\usepackage{babel}
\begin{document}
\title{Temporal Wheeler\textquoteright s delayed-Choice Experiment based
on Cold Atomic Quantum Memory}
\author{Ming-Xin Dong}
\affiliation{Key Laboratory of Quantum Information, University of Science and Technology
of China, Hefei, Anhui 230026, China.}
\affiliation{Synergetic Innovation Center of Quantum Information and Quantum Physics,
University of Science and Technology of China, Hefei, Anhui 230026,
China.}
\author{Dong-Sheng Ding}
\email{dds@ustc.edu.cn}

\affiliation{Key Laboratory of Quantum Information, University of Science and Technology
of China, Hefei, Anhui 230026, China.}
\affiliation{Synergetic Innovation Center of Quantum Information and Quantum Physics,
University of Science and Technology of China, Hefei, Anhui 230026,
China.}
\author{Yi-Chen Yu}
\affiliation{Key Laboratory of Quantum Information, University of Science and Technology
of China, Hefei, Anhui 230026, China.}
\affiliation{Synergetic Innovation Center of Quantum Information and Quantum Physics,
University of Science and Technology of China, Hefei, Anhui 230026,
China.}
\author{Ying-Hao Ye}
\affiliation{Key Laboratory of Quantum Information, University of Science and Technology
of China, Hefei, Anhui 230026, China.}
\affiliation{Synergetic Innovation Center of Quantum Information and Quantum Physics,
University of Science and Technology of China, Hefei, Anhui 230026,
China.}
\author{Wei-Hang Zhang}
\affiliation{Key Laboratory of Quantum Information, University of Science and Technology
of China, Hefei, Anhui 230026, China.}
\affiliation{Synergetic Innovation Center of Quantum Information and Quantum Physics,
University of Science and Technology of China, Hefei, Anhui 230026,
China.}
\author{En-Ze Li}
\affiliation{Key Laboratory of Quantum Information, University of Science and Technology
of China, Hefei, Anhui 230026, China.}
\affiliation{Synergetic Innovation Center of Quantum Information and Quantum Physics,
University of Science and Technology of China, Hefei, Anhui 230026,
China.}
\author{Lei Zeng}
\affiliation{Key Laboratory of Quantum Information, University of Science and Technology
of China, Hefei, Anhui 230026, China.}
\affiliation{Synergetic Innovation Center of Quantum Information and Quantum Physics,
University of Science and Technology of China, Hefei, Anhui 230026,
China.}
\author{Kan Zhang}
\affiliation{Department of Fundamental Education, Anhui Institute of Information
Technology, Wuhu, Anhui 241002, China.}
\author{Da-Chuang Li}
\affiliation{Institute for Quantum Control and Quantum Information and School of
Physics and Materials Engineering, Hefei Normal University, Hefei,
Anhui 230601, Peolpe's Republic of China.}
\author{Guang-Can Guo}
\affiliation{Key Laboratory of Quantum Information, University of Science and Technology
of China, Hefei, Anhui 230026, China.}
\affiliation{Synergetic Innovation Center of Quantum Information and Quantum Physics,
University of Science and Technology of China, Hefei, Anhui 230026,
China.}
\author{Bao-Sen Shi}
\email{drshi@ustc.edu.cn}

\affiliation{Key Laboratory of Quantum Information, University of Science and Technology
of China, Hefei, Anhui 230026, China.}
\affiliation{Synergetic Innovation Center of Quantum Information and Quantum Physics,
University of Science and Technology of China, Hefei, Anhui 230026,
China.}
\date{\today}

\maketitle
\textbf{Nowadays the most intriguing features of wave-particle complementarity
of single photon is exemplified by the famous Wheeler's delayed choice
experiment in linear optics, nuclear magnetic resonance and integrated
photonic device systems. Studying the wave-particle behavior in light
and matter interaction at single photon level is challenging and interesting,
which gives how single photons complement in light and matter interaction.
Here, we demonstrate a Wheeler\textquoteright s delayed choice experiment
in an interface of light and atomic memory, in which the cold atomic
memory makes the heralded single photon divided into a superposition
of atomic collective excitation and leaked pulse, thus acting as memory
beam-splitters. We observe the morphing behavior between particle
and wave of a heralded single photon by changing the relative proportion
of quantum random number generator, the second memory efficiency,
and the relative storage time of two memories. The reported results
exhibit the complementarity behavior of single photon under the interface
of light-atom interaction. }

The wave-particle duality or complementarity \cite{greiner2011quantum}
in quantum physics has been demonstrated by Wheeler\textquoteright s
delayed-choice experiment \cite{wheeler1978past,hellmuth1987delayed,baldzuhn1989wave,lawson1996delayed,kim2000delayed,jacques2007experimental,jacques2008delayed}
exhibiting its paradoxical nature, in which a photon is forced to
choose a behavior before the observer decides what to measure \cite{ma2016delayed}.
This wave-particle duality is the heart of quantum mechanics, because
it is introduced to understand intuitively the behavior of quantum
particles. Single photon's wave-particle duality are studied in many
systems of linear optics \cite{tang2012realization,kaiser2012entanglement,jacques2007experimental}
and integrated photonic device \cite{peruzzo2012quantum}, in these
experiments a Mach-Zender interferometer is configured in which a
photon passing through it exhibits wave- or particle-like features
depending on the experimental apparatus it is confronted by. By a
proposal of using a 'quantum' beam splitter (BS) \cite{ionicioiu2011proposal},
people can investigate the intermediate behaviour between wave and
particle nature \cite{tang2012realization,kaiser2012entanglement,peruzzo2012quantum}.
Moreover, the delayed-choice quantum erasure \cite{scully1982quantum,Kim2000}
and quantum entanglement swapping \cite{jennewein2001experimental,ma2012experimental}
are reported with various physical systems. Although the fundamental
aspects of the delayed-choice experiments have been well studied in
these systems, the delayed-choice experiment under the picture of
light-matter interaction was rather obscure.

Light interaction with matter offers a rich of physics \cite{scully1999quantum},
such as photon absorption, spontaneous emission and photon storage
and so on. One of interesting phenomena is quantum memory, a device
that can coherently store and retrieve single photon including its
information \cite{kozhekin2000quantum}. An intriguing question may
arise as to what happens to wave-particle duality when light interact
with matter. When a atom absorbs a single photon with less than one
hundred percent, the question is whether the photon is leaked or absorbed
by the atom is dependent on the choice of the observer.

In this work, we demonstrate a Wheeler\textquoteright s delayed-choice
experiment based on atomic quantum memory. Here, three cold $^{85}$Rb
atomic ensembles trapped in a 2D magneto-optical trap are utilized,
in which one ensemble is used to generate a heralded single photon
and the other two act as the temporal beam splitters based on Raman
storage protocol \cite{nunn2007mapping,reim2010towards,reim2012multipulse,ding2015raman}
configuring a temporal Mach-Zender interferometer. The memory we used
here acts as a quantum device that divides the single photon packet
into atomic and photonic components when the memory efficiency is
less than unitary. We observe the morphing phenomenon between particle
and wave behavior by changing a serial of experimental parameters.
Our reported results give an important viewpoint that the single photon
has a non-locality property under the interface of light and matter
interaction.

\begin{figure*}
\includegraphics[width=1.95\columnwidth]{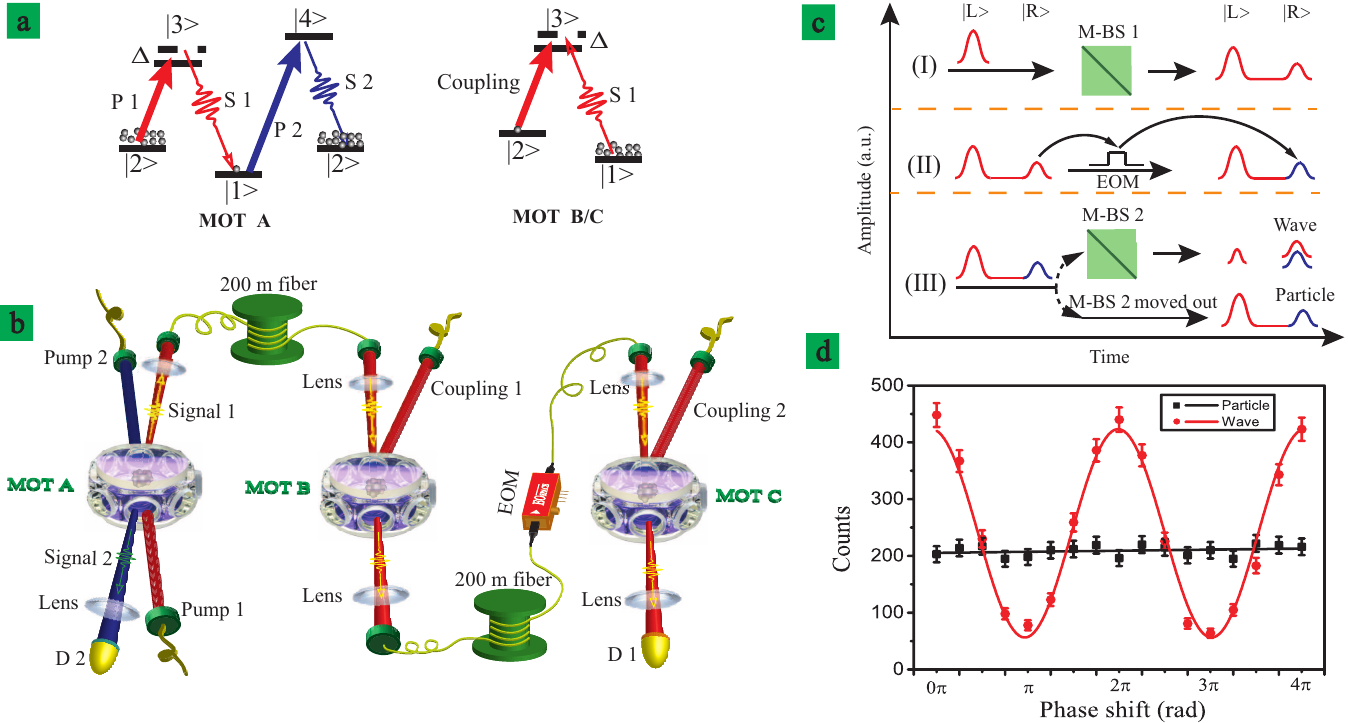}\caption{\textbf{Experimental realization of Wheeler's delayed-choice experiment.}
(a) Energy level diagram. MOT A, B and C represent three magneto-optical
traps. Single photons S 1 and S 2 are generated from MOT A using SFWM
process, and the MOT B/C acting as quantum memory is based on Raman
storage protocol with single-photon detuning $\Delta/2\pi=50$ MHz.
States \textbar 1\textgreater , \textbar 2\textgreater , \textbar 3\textgreater{}
and \textbar 4\textgreater{} correspond to $^{85}$Rb atomic levels
of $5S_{1/2}(F=2)$, $5S_{1/2}(F=3)$, $5P_{1\text{/2}}(F=3)$ and
$5P_{3/2}(F=3)$ respectively. (b) Simplified experimental setup.
Pump 1/2 is pump light beam, Coupling 1/2 represents the coupling
light beam. EOM is electro-optic modulator, introducing a phase shift,
and 200 m fiber is used for optical delay of 1 $\mu s$. D 1/D 2,
avalanche diode 1/2. (c) A simplified sketch of timing sequence in
delayed-choice experiment. (d) The wave-particle duality of single-photon.
When the M-BS 2 is inserted, the red experimental data dots are obtained
and interference curve is fitted with a sine function, which is a
wavelike phenomenon. The black line are fitted with a constant function
while M-BS 2 is moved out, in which the interference is vanished,
revealing a particle-like phenomenon. }

\label{setup}
\end{figure*}
The simple energy level diagram is shown in Fig. \ref{setup}(a).
We firstly generate a Stokes and anti-Stokes photon (Signal 2 and
Signal 1) through spontaneously four-wave mixing (SFWM) \cite{balic2005generation,du2008narrowband}
process in MOT A, as shown in Fig. \ref{setup}(b). Here, pump 1 (795
nm, Rabi frequency $2\pi\times1.19$ MHz) and pump 2 (780 nm, Rabi
frequency $2\pi\times14.79$ MHz) are orthogonal polarization and
propagate counter collinearly in MOT A with optical depth (OD) of
40. The angle between pump lasers and signals is $2.8^{\circ}$, and
we collect the signal 2 (780 nm) and signal 1 (795 nm) by using lens
with focal length of 300 mm. Since signal 2 is detected the by single
photon counting module (avalanche diode 2, PerkinElmer SPCM-AQR-15-FC,
maximum dark count rate of 50/s), a heralded single photon signal
1 is obtained and then coupled into a 200-m single-mode fiber.

Quantum memory can be served as a quantum device that makes the photon
pulse separate in timeline, the separated time interval and the amplitude
can be arbitrarily configured, thus called as a dynamically configurable
temporal BS \cite{reim2012multipulse}. Here we exploit two Raman
memories MOT B and MOT C as memory-based beam splitters (M-BSs) to
configure an interferometer in temporal domain. As depicted in Fig.
\ref{setup}(c), the implement of delayed-choice scheme is illustrated
as follow. The Signal 1 photon is split into two parts by M-BS 1 with
the following expressed state

\begin{equation}
\left|{\rm \mathit{\psi}}\right\rangle =\sqrt{1-\eta_{1con}}\left|{\rm \mathit{L}}\right\rangle +e^{i\theta_{1}}\sqrt{\eta_{1con}}\left|{\rm \mathit{R_{a\mathrm{1}}}}\right\rangle
\end{equation}
here, $\eta_{1con}$, is the conversion efficiency of optical signal
to spin wave in MOT B. The right two terms in above equation represent
the split states corresponding to the leaked part $\left|{\rm \mathit{L}}\right\rangle $
and stored part $\left|{\rm \mathit{R_{a\mathrm{1}}}}\right\rangle $
under the quantum memory process respectively, the coefficients $\sqrt{1-\eta_{1con}}$
and $\sqrt{\eta_{1con}}$ are the amplitude of these two parts. $\theta_{1}=w\cdot\triangle t$
is the relative phase between the states $\left|{\rm \mathit{L}}\right\rangle $
and $\left|{\rm \mathit{R_{a\mathrm{1}}}}\right\rangle $ with the
storage time $\triangle t$. The stored part $\left|{\rm \mathit{R_{a\mathrm{1}}}}\right\rangle $
corresponds to the atomic collective excited state defined in Methods.
The expression given by Eq. (1) corresponds to superposition state
of photon and atom, thus we don't know whether the photon is transformed
to atomic state or is leaked.

\begin{figure*}
\includegraphics[width=1.95\columnwidth]{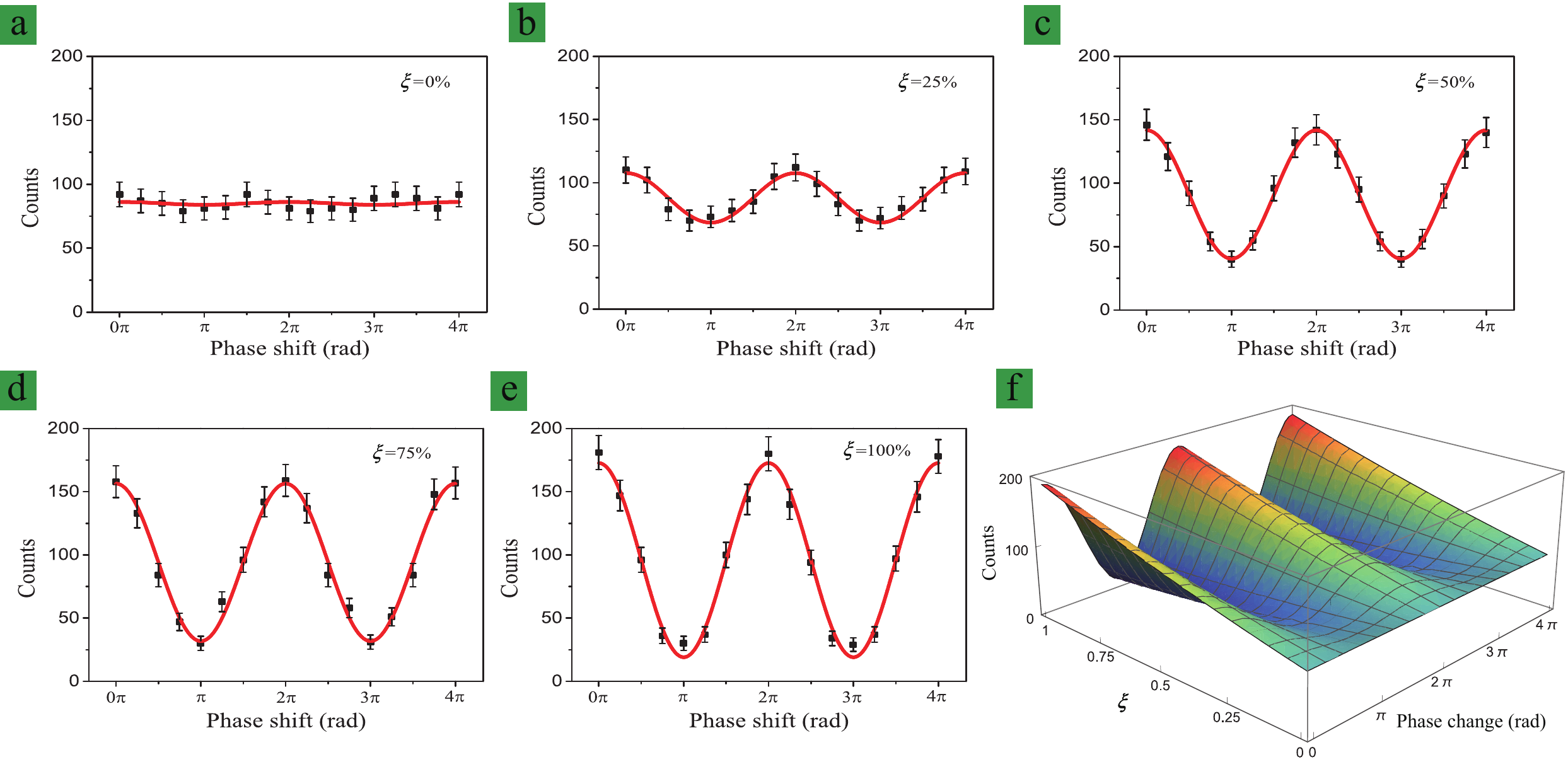}\caption{\textbf{Morphing phenomenon between particle and wave behavior.} (a)-(e)
The recorded coincidence counts against varying the EOM phase in a
step of $\pi/4$. The red curves are fitted $N\mid\xi\sqrt{1-\eta_{1con}}\sqrt{\eta_{2}}+\sqrt{\eta_{1}}e^{i\varphi_{EOM}}\mid^{2}$,
$N$ is the total photon counts $N=611\pm32$, $\eta_{1}=0.133\pm0.004,$
$\eta_{1con}=0.850\pm0.026$, $\eta_{2}=0.24$. The fitted $\xi$
is $0.01$, $0.24$, $0.53$, $0.74$, $0.96$ from (a)-(e) respectively.
(f) Simulated result of continuous morphing from wave to particle
behavior with $N=611$, $\eta_{1}=0.133,$ $\eta_{1con}=0.850$, $\eta_{2}=0.24$.}

\label{changing R}
\end{figure*}
After $\triangle t=200$ ns storage time, we turn on the coupling
laser to read the spin wave in MOT B out as $\left|{\rm \mathit{R}}\right\rangle $.
Between MOT B and MOT C, we make signal travel through a 200-m fiber
for optical delay of about 1 $\mu s$ to enlarge the coherence length
of interferometer (see Methods), the signal 1 photon has a photonic
superposition $\left|{\rm \mathit{\psi_{\mathrm{1}}}}\right\rangle \sim\sqrt{1-\eta_{1con}}\left|{\rm \mathit{L}}\right\rangle +\sqrt{\eta_{1}}e^{i\theta_{1}}\left|{\rm \mathit{R}}\right\rangle $.
(Here, $\eta_{1}=\eta_{1con}\eta_{1stored}$, is the total storage
efficiency of optical signal in MOT B, including the efficiency of
optical signal conversion to spin wave and spin wave retrieval to
optical excitation $\eta_{1stored}$.) These two split photon packets
distinct in time domain are equivalent to the two arms of interferometer.
We vary the relative phase between two interferometer arms by modulating
a phase shift on $\left|{\rm \mathit{R}}\right\rangle $ by an electro-optical
modulation (EOM), thus the state becomes

\begin{equation}
\left|{\rm \mathit{\psi}}\right\rangle \sim\sqrt{1-\eta_{1con}}\left|{\rm \mathit{L}}\right\rangle +\sqrt{\eta_{1}}e^{i\theta_{1}+\varphi_{EOM}}\left|{\rm \mathit{R}}\right\rangle
\end{equation}
here, $\varphi_{EOM}$ is the added phase by EOM.

As soon as the leaked part arrives at M-BS 2 (in MOT C), a randomly
choice to insert or remove the M-BS 2 has been made to realize the
Wheeler's delayed choice experiment, which is controlled by a quantum
random number generator (QRNG) with a density matrix $\rho=(1-\xi)\left|0\right\rangle _{s2}\left\langle 0\right|_{s2}+\xi\left|1\right\rangle _{s2}\left\langle 1\right|_{s2}$,
the switch on or off for coupling 2 depends on states $\left|0\right\rangle _{s2}$
and $\left|1\right\rangle _{s2}$. If M-BS 2 is removed/inserted ($\xi=0/1$),
the whole setup forms a open/closed Mach-Zender interferometer, the
leaked part is not-converted/converted to the spin wave in MOT C by
switching off the coupling 2 light with Rabi frequency $\Omega_{c2}=2\pi\times24.21$
MHz. The state is written as

\begin{equation}
\begin{split}\left|{\rm \mathit{\psi}}\right\rangle  & \sim\sqrt{1-\eta_{1con}}\xi(\sqrt{1-\eta_{2con}}\left|{\rm \mathit{L}}\right\rangle +\sqrt{\eta_{2con}}e^{i\theta_{2}}\left|{\rm \mathit{{\rm \mathit{R_{a\mathrm{2}}}}}}\right\rangle )\\
 & +\sqrt{1-\eta_{1con}}(1-\xi)\left|{\rm \mathit{L}}\right\rangle +\sqrt{\eta_{1}}e^{i\theta_{1}+\varphi_{EOM}}\left|{\rm \mathit{R}}\right\rangle
\end{split}
\end{equation}

here, $\eta_{2con}$ is the conversion efficiency of leaked part $\left|{\rm \mathit{L}}\right\rangle $
in Eq. (2) to spin wave in MOT C. After the same storage time of 200
ns, thus the relative phase $\theta_{1}=\theta_{2}=\triangle\theta$,
we retrieve the spin wave to optical signal by considering the efficiency
$\eta_{2stored}$ of spin wave retrieval to optical excitation,

\begin{equation}
\begin{split}\left|{\rm \mathit{\psi}}\right\rangle  & \sim(\sqrt{1-\eta_{1con}}\xi\sqrt{1-\eta_{2con}}+\sqrt{1-\eta_{1con}}(1-\xi))\left|{\rm \mathit{L}}\right\rangle \\
 & +e^{i\triangle\theta}(\xi\sqrt{1-\eta_{1con}}\sqrt{\eta_{2}}\left|{\rm \mathit{{\rm \mathit{R}}}}\right\rangle +\sqrt{\eta_{1}}e^{i\varphi_{EOM}}\left|{\rm \mathit{R}}\right\rangle )
\end{split}
\end{equation}

here, $\eta_{2}=\eta_{2con}\eta_{2stored}$ is the total storage efficiency
of optical signal in MOT C, including the efficiency of optical signal
conversion to spin wave and spin wave retrieval to optical excitation.
We can check the photon interference by detecting the retrieved part
$\left|{\rm \mathit{R}}\right\rangle $, it osculates with a function
of

\begin{equation}
P(\eta_{1},\eta_{2},\xi,\varphi_{EOM})\sim\mid\xi\sqrt{1-\eta_{1con}}\sqrt{\eta_{2}}+\sqrt{\eta_{1}}e^{i\varphi_{EOM}}\mid^{2}
\end{equation}

In the first case, if M-BS 2 is inserted (for $\xi=1$), the two arms
of interferometer are recombined and we can observe an wave-like phenomenon
sketched in the red curve in Fig. \ref{setup}(d). In the second case,
while the M-BS2 is removed ($\xi=0$), the interferometer remains
open, and we observe no interference, revealing the particle nature
of photon, as shown in the black line in Fig. \ref{setup}(d). In
the second case, the open interferometer corresponds to the situation
that the leaked part is not converted to spin wave and passes through
the MOT C directly, in which we observe no interference because there
is no overlap of split signals.

\begin{figure*}
\includegraphics[width=1.95\columnwidth]{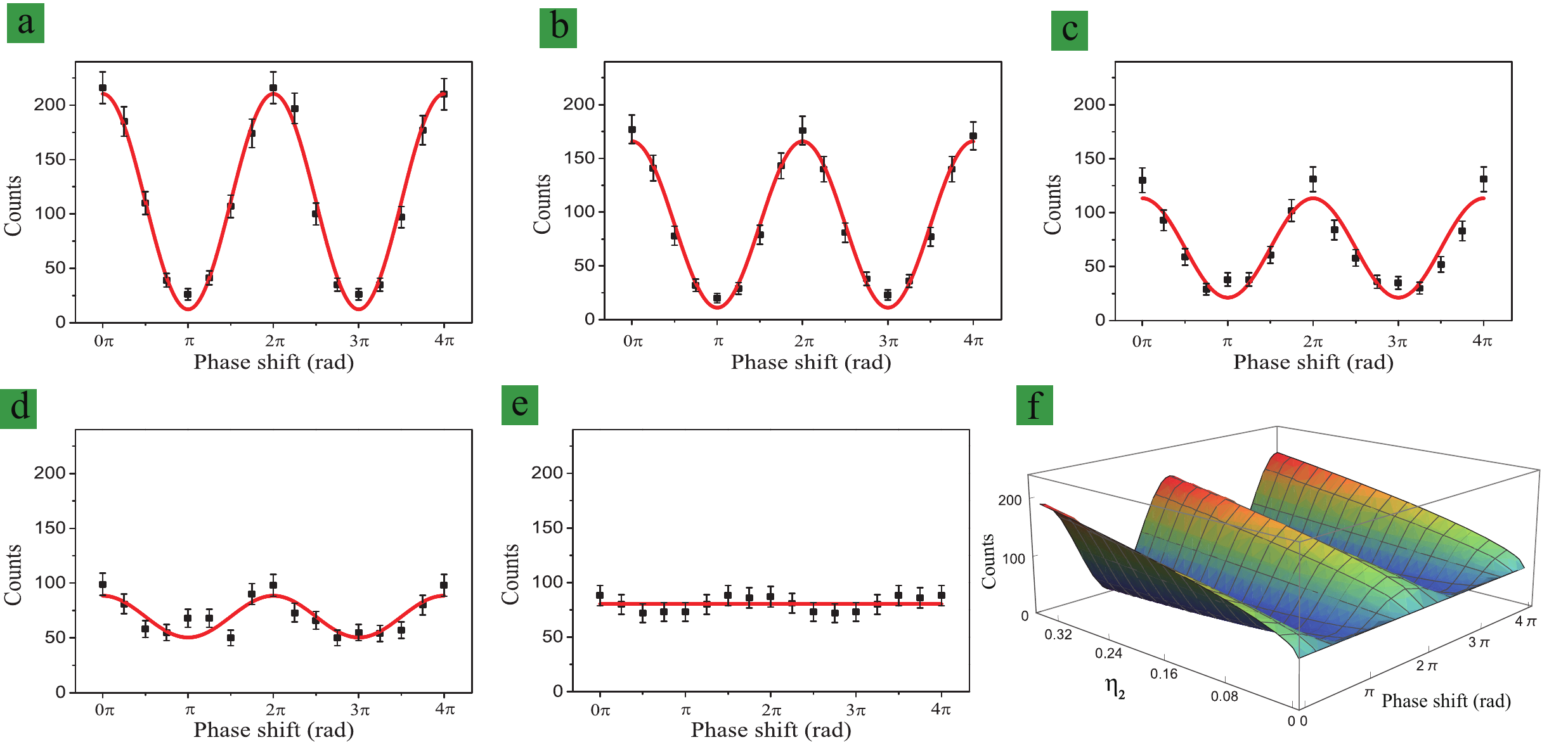}\caption{\textbf{Demonstration of interference ability of our interferometer}.
(a)-(e) The interference pattern with different storage efficiency
of spin wave in MOT C. The red curves are fitted $N\mid\sqrt{1-\eta_{1con}}\sqrt{\eta_{2}}+\sqrt{\eta_{1}}e^{i\varphi_{EOM}}\mid^{2}$
(where $N=568\pm40$, $\eta_{1}=0.122\pm0.011,$ $\eta_{1con}=0.850,$
from (a)-(e), $\eta_{2}=0.331$, $0.259,$ $0.114,$ $0.015,$ $0$
respectively). (f) The simulated interference ability of interferometer
with the change of $\eta_{2}$ (where we set $N=568$, $\eta_{1}=0.122,$
$\eta_{1con}=0.850$). }

\label{against efficiency}
\end{figure*}
We demonstrate a morphing phenomenon between wave- and particle-like
behavior by changing the relative proportion of QRNG $\xi=0$, $25
$, $50\%$, $75\%$, $1$. Generally, a quantum delayed-choice experiments
require an ancilla which is prepared in a superposition state $\left|\varPsi\right\rangle =\cos\alpha\left|0\right\rangle +\sin\alpha\left|1\right\rangle $
and then measured, the results controls the insert or remove of BS
\cite{tang2012realization,peruzzo2012quantum}, the morphing between
wavelike and particle-like behavior is observed. In our scheme, the
ancilla can be expressed as a mixed state $\rho=(1-\xi)\left|0\right\rangle \left\langle 0\right|+\xi\left|1\right\rangle \left\langle 1\right|$,
here $(1-\xi$) is the probability of the vacuum state. In Fig. \ref{changing R},
$\xi$ takes different values and we observe a phenomenon from particle
to wave. The calculated visibility of interference is $P(\eta_{1},\eta_{2},\xi,\varphi_{EOM})$,
and the measured visibility is not very high, this is caused by the
mismatching of two retrieved signals because the bandwidth of two
memories is slightly different from each other.

In an analog to operational definition given in \cite{ionicioiu2011proposal},
the ``ability'' or ``inability'' to generate interference can
be utilized to describe the wave or particle properties. In the following
part, we explore the relationship between interference ability of
our apparatus and storage parameters such as storage efficiency and
storage time, which is in favor of understanding wave-particle complementarity
in the light-matter interaction. As illustrated in Fig. \ref{against efficiency},
the visibility is varied against different storage efficiencies in
MOT C by varying the Rabi frequency of coupling 2 light from $2\pi\times27.86$
to $\text{0}$. The maximum visibility (Fig. \ref{against efficiency}(a))
corresponds to the storage efficiency of $33.1{\rm \%}$ in MOT C.
Here, in order to obtain the perfect interference, we should balance
the retrieved signals after two storage processes. Therefore, we choose
a suitable storage efficiency of spin wave in MOT C by varying the
Rabi frequency of coupling 2. In addition, it is also crucial to the
choose a suitable storage efficiency in MOT B. Because if the storage
efficiency in MOT B is too large, the leaked part as the input of
the second storage process is too little to obtain the enough retrieved
signal after leaving out of MOT C. As a result, in our experiment
we actually optimize the storage efficiencies in MOT B and MOT C to
achieve the best interference. The minimum visibility (Fig. \ref{against efficiency}(e))
corresponds to the MOT C storage efficiency of 0, revealing the nature
of particle. Fig. \ref{against efficiency}(f) is the simulated interference
against the effective storage efficiency of memory in MOT C.

\begin{figure*}
\includegraphics[width=1.95\columnwidth]{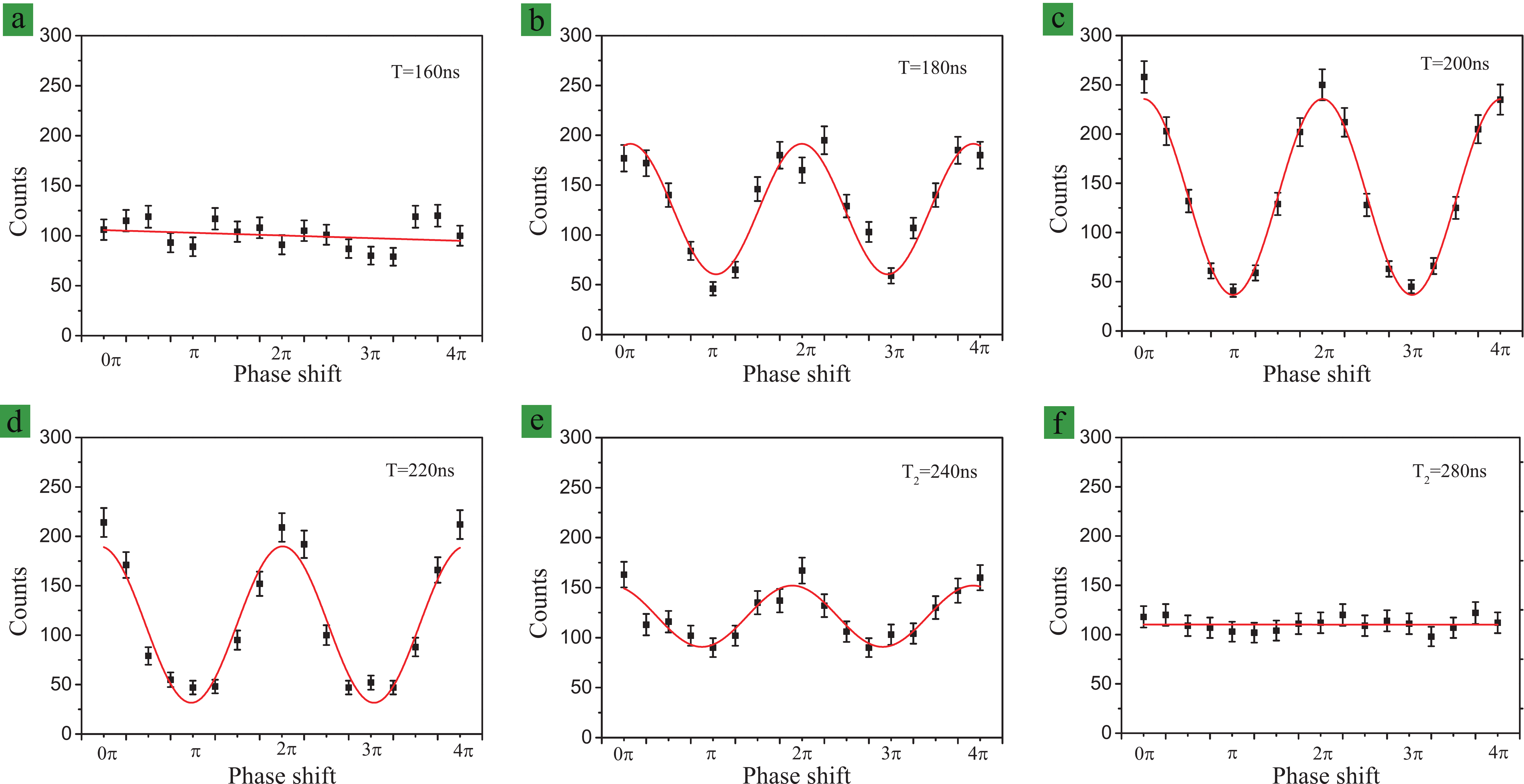}\caption{\textbf{The demonstration of quantum memory acting as the temporal
BS.} (a)-(f) The interference phenomenon with different storage time
of spin wave in MOT C from T=160 ns to 280 ns. Dots are experimental
data and red curves are fitted sine function or constant function. }

\label{changing time}
\end{figure*}
In our scheme, it is intriguing to study the wave-particle complementarity
using our controllable M-BS based on quantum memory. As a result,
we attempt to vary the storage time of spin wave in MOT C and observe
a morphing from wave to particle nature sketched in Fig. \ref{changing time}.
The best interference pattern is shown in Fig. \ref{changing time}(c)
with the storage time of 200 ns, which is identical to the storage
time in MOT B. Intrinsically, the visibility of interference is positive
correlated to the overlap among two retrieval signals. While we vary
the storage time of spin wave in MOT C, the degree of two parts of
signal 1 overlap in temporal domain is also changing. There is almost
no interference pattern in Fig. \ref{changing time}(a/f) with the
storage time of 160/280 ns, in which the two retrieved signals are
almost separated with no overlap. The overlap time window is $\sim$
120 ns ($=280-160$ ns), which is approach to the coherence time of
retrieved optical mode of 110 ns. In addition, the overlap of two
retrieved signals can also controlled by adjusting the waveforms or
bandwidth of two retrieval wave packets.

\begin{figure*}[t]
\includegraphics[width=1.95\columnwidth]{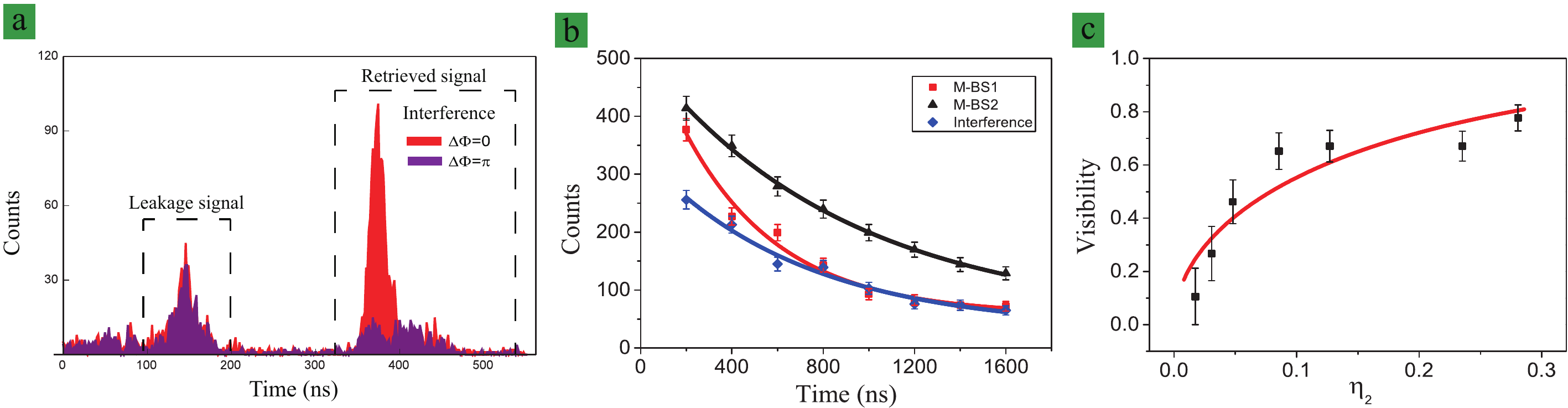}\caption{\textbf{The illustration of decoherence in our system.} (a) The interference
of two spin waves stored in MOT B and MOT C with the modulated phase
of 0 and $\pi$ by EOM, corresponding to the red and blue parts respectively.
(b) The recorded coincidence counts against storage time in MOT B
(red), and MOT C (black). The red and black curves are fitted $Ae^{-t/T1}+g_{0}$
(where red curve, $A=503,$ $T1=420,$$g_{0}=58,\text{and back curve,}$
$A=457,$ $T1=893,$ $g_{0}=51$). The interference results (blue)
with modulated phase of EOM $\Delta\Phi=0$ against the storage time,
here the storage time in MOT B and MOT C are set identically. The
blue curve is also fitted $Ae^{-t/T2}+g_{0}$ (where $A=304,$ $T2=691,$
$g_{0}=32$). (c) The interference visibility against the OD of atomic
ensemble in MOT C. The red curve corresponds to theoretical fitting
(where $\eta_{1}=0.132,$ $\eta_{1con}=0.88$). }

\label{decoherence}
\end{figure*}

In Fig. \ref{decoherence}(a), the recorded coincidence counts with
$\Delta\Phi=0$ and $\Delta\Phi=\pi$ are illustrated in red and purple
data respectively. The demonstration of wave and particle duality
is not only dependent on the storage efficiencies $\eta_{1}$ and
$\eta_{2}$, but also the coherence of the M-BSs. To show the decoherence
in our system, we explore the decoherence of two memories and interferometer
respectively as shown in Fig. \ref{decoherence}(b). The red and black
curves describe the coherence of M-BS 1 in MOT B and M-BS 2 in MOT
C with coherence time of 420 ns and 893 ns respectively. The bule
curve is the coherence (with coherence time 691 ns) of interferometer
when the phase of EOM $\Delta\Phi=0$. Ultimately, we explore the
relationship between OD of MOT C and interference visibility. Intrinsically,
the change of OD of MOT C corresponds to the variation of storage
efficiency in MOT C $\eta_{2}$. We vary the OD of MOT C from 0 to
40, meanwhile we measure the $\eta_{2}$. The visibility can be expressed
as $V=[P_{max}-P_{min}${]}/$[P_{max}+P_{min}${]}, a function of
$\eta_{2}$, and we fit it with our experimental data as shown in
Fig. \ref{decoherence}(c).

Our Wheeler\textquoteright s delayed-choice experiment rely on the
key element of M-BS, its properties can be turned arbitrarily which
are very different from other's demonstrations \cite{tang2012realization,peruzzo2012quantum,kaiser2012entanglement}.
For example, the wave-particle duality demonstrated here is dependent
on the pulse matching, the amplitude between two retrieved signals
as described in above. The bandwidth and amplitude of two retrieved
signals can be turned by changing the OD in MOT B and MOT C and the
Rabi frequencies $\Omega_{c1}$ and $\Omega_{c2}$. In addition, the
M-BS used here in principle can configure high-dimensional Mach-Zender
interferometer with multiple temporal arms by addressing multipulse
with a Raman quantum memory \cite{reim2012multipulse}, by which,
one can demonstrate the Wheeler\textquoteright s delayed-choice experiment
with multiple photonic paths.

The QRNG used here is from the photon pair generated from SFWM process,
which is a mixed state $\rho=(1-\xi)\left|0\right\rangle _{s2}\left\langle 0\right|_{s2}+\xi\left|1\right\rangle _{s2}\left\langle 1\right|_{s2}$,
not from a superposition state \cite{tang2012realization,peruzzo2012quantum,kaiser2012entanglement}
or classical choices \cite{hellmuth1987delayed,baldzuhn1989wave,lawson1996delayed,kim2000delayed,jacques2007experimental,jacques2008delayed}.
So the morphing phenomenon we measured correspond to the intermediate
between quantum and classical situations. The intriguing physics here
we want to emphasize is the complementarity of single photon interacted
with quantum memory, exhibiting an important relationship between
single photon and the atoms under interaction. The non-locality reported
here is not only the two photonic temporal arms but also the states
between atomic spin wave and the leaked signal.

In summary, we have demonstrated a Wheeler\textquoteright s delayed-choice
experiment with Raman memory temporal beam splitters in two atomic
ensembles, which construct a temporal Mach-Zender interferometer with
a 200 meter fiber. The wave- and particle-like morphing behavior of
heralded single photon is demonstrated by changing the experimental
parameters of relative proportion, the storage efficiency, the optical
depth and coherence of the M-BSs. The resulting Wheeler\textquoteright s
delayed-choice experiment under light-atom interaction gives a fundamental
aspect that the single photon exhibits a non-locality when interacting
with atoms.

\section*{Method Sections}

\textbf{Experimental time sequence.} The repetition rate of our experiment
is 100 Hz, and the MOT trapping time is 8.7 ms. Moreover, the experimental
window is 1.3 ms. The fields of pumps 1 and 2 are controlled by two
acousto-optic modulators (AOMs) modulated by arbitrary function generator
(Tektronix, AFG3252). Two lenses L1 and L2, each with a focal length
of 300 mm, are used to couple the signal fields into the atomic ensemble
in MOT 1. The fields of pumps 1 and 2 are collinear, and hence their
respective signal fields are collinear. The vector matching condition
$k_{p1}-k_{S1}=k_{p2}-k_{S2}$ is satisfied in the spontaneous four-wave
mixing process, as the methods are the same as in our previous work.
The two signal photons are collected into their respective single-mode
fibers and are detected by two single photon detectors (avalanche
diode, PerkinElmer SPCM-AQR-16-FC, $60
$ efficiency, maximum dark count rate of 25/s). The two detectors are
gated in the experimental window. The gated signals from the two detectors
are then sent to a time-correlated single photon counting system (TimeHarp
260) to measure their time-correlated function.

\textbf{Raman quantum memory}. Two atomic ensembles in MOT B and MOT
C are serving as Raman memories \cite{ding2015raman}. The specific
storage process is illustrated as follow: The signal 1 photon is directed
through the MOT B with OD of 35, and simultaneously we adiabatically
switch off the coupling 1 light with Rabi frequency $\Omega_{c1}=2\pi\times20.61$
MHz and a beam waist of 2 mm, and then a stored atomic collective
excitation is obtained given by $1/\sqrt{m}\sum e^{i\mathbf{\mathit{k}}_{S}\mathbf{\cdot\mathit{r}}_{i}}\left|1\right\rangle _{1}\cdot\cdot\cdot\left|2\right\rangle _{i}\cdot\cdot\cdot\left|1\right\rangle _{m}$
\cite{fleischhauer2000dark}, also called as spin wave. $\mathbf{\mathrm{\mathit{k}}}_{S}=\mathbf{\mathrm{\mathit{k}}}_{c1}-\mathbf{\mathrm{\mathit{k}}}_{s1}$
is the wave vector of atomic spin wave, $\mathbf{\mathrm{\mathit{k}}}_{c1}$
and $\mathbf{\mathit{k}}_{s1}$ are the vectors of coupling and signal
1 fields, $\mathbf{\mathit{r}}_{i}$ denotes the position of the $i$-th
atom in atomic ensemble in MOT B. After a programmable storage time,
the spin wave is converted back into photonic excitation by switching
on the coupling 1 light again. Due to the Raman memory efficiency
is significantly dependent on the OD of atoms \cite{guo2019high},
the input single photon would induces a leaked component in the storage
process with a controllable OD.

\textbf{The coherence of Mach-Zender interferometer.} The two signal
parts$\text{\ensuremath{\left|{\rm \mathit{L}}\right\rangle },}$
$\left|{\rm \mathit{R}}\right\rangle $ (leaked and retrieved signals)
marked in Fig.\ref{setup} (c) are exploited to construct the two
arms of the temporal Mach-Zender interferometer. We enlarge the length
of interferometer arms by inserting a 200-m optical fiber (corresponding
to a time delay of $\sim$1 $\mu s$) in signal 1's optical path to
avoid the retrieved signal interferes with itself on M-BS 2 if the
length of interferometer arm is shorter than the coherence length
of $\left|{\rm \mathit{R}}\right\rangle $. The coherence length of
$\left|{\rm \mathit{R}}\right\rangle $ is determined by the time
width of the signal 1 photon, which is 50 ns for our experiment.

\textbf{Quantum random number generator.} We generate the random number
by performing a logic gate operation between the detection of Stokes
photon and a 100 kHz transistor-transistor logic (TTL) signal generated
by an arbitrary function generator. Intrinsically, the emission of
single photon is a SFWM process, the generated QRNG is described by
the operator $\rho=(1-\xi)\left|0\right\rangle _{s2}\left\langle 0\right|_{s2}+\xi\left|1\right\rangle _{s2}\left\langle 1\right|_{s2}$,
the coefficient $\xi$ can be adjusted by the duty cycle of TTL signal,
the first term $(1-\xi)\left|0\right\rangle _{s2}\left\langle 0\right|_{s2}$
is used to switch the coupling 2 laser off and the second term $\xi\left|1\right\rangle _{s2}\left\langle 1\right|_{s2}$
to switch the coupling 2 laser on. We measure a morphing phenomenon
of photon that behaves from wave to particle by changing the duty
cycle of TTL signal.

\textbf{Author contributions }Ming-Xin Dong and Dong-Sheng Ding contribute
to this paper equally. D.S.D. conceived the idea and experiment. M.X.D.
designed and carried out the experiments with assistance from W.H.Z.
and Y.C.Y.. M.X.D. wrote the manuscript with contributions from B.S.S..
D.S.D., B.S.S. and G.C.G. supervised the project.

\textbf{Competing financial interests} The authors declare no competing
financial interests.

\textbf{Acknowledgments} This work was supported by National Key R\&D
Program of China (2017YFA0304800), the National Natural Science Foundation
of China (Grant Nos. 61525504, 61722510, 61435011, 11174271, 61275115,
11604322), and the Innovation Fund from CAS, Anhui Initiative in Quantum
Information Technologies (AHY020200).

\bibliographystyle{apsrev4-1}

\end{document}